\documentclass[12pt,usenames,dvipsnames]{article}

\usepackage{latexsym}
\usepackage{amssymb,amsfonts,amsmath}
\usepackage{graphicx} 
\usepackage{indentfirst}
\usepackage{bbm}
\usepackage{amssymb}
\usepackage{verbatim}
\usepackage{amsmath, amsthm,amssymb}
\usepackage{mathrsfs}
\usepackage{hyperref}
\usepackage{amsfonts}
\usepackage{dsfont}
\usepackage{cite}
\usepackage{xcolor}
\usepackage{enumerate}
\usepackage{cleveref}

\parskip=\medskipamount

\newcommand {\cD}{{\cal D}}

\newcommand {\cG}{{\cal G}}
\newcommand {\cH}{{\cal H}}

\newcommand {\cJ}{{\cal J}}

\newcommand {\cN}{{\cal N}}
\newcommand {\cO}{{\cal O}}

\newcommand {\cT}{{\cal T}}

\newcommand {\cV}{{\cal V}}
\newcommand {\cW}{{\cal W}}

\newcommand {\cY}{{\cal Y}}

\def\a{\alpha}

\def\b{\beta}

\def\d{\delta}

\def\f{\phi}
\def\g{\gamma}
\def\G{\Gamma}

\def\j{\psi}

\def\p{\pi}
\def\q{\theta}

\def\s{\sigma}

\def\x{\xi}
\def\z{\zeta}
\def\D{\Delta}
\def\F{\Phi}
\def\J{\Psi}

\def\O{\Omega}

\def\U{\Upsilon}

\def\rd{{\rm d}}
\def\ri{{\rm i}}

\def\N{{\cal N}}

\newcommand{\ad}{{\dot{\alpha}}}                           
\newcommand{\bd}{{\dot{\beta}}}                            
\newcommand{\ve}{\varepsilon}                            

\renewcommand{\aa}{{\a\ad}}
\newcommand{\bb}{{\b\bd}}
\newcommand{\pa}{\partial}                           
\newcommand{\hf}{\frac12}

\newcommand{\be}{\begin{equation}}
\newcommand{\ee}{\end{equation}}
\newcommand{\bea}{\begin{eqnarray}}
\newcommand{\eea}{\end{eqnarray}}
\newcommand{\non}{\nonumber}
\newcommand{\1}{{\underline{1}}}
\newcommand{\2}{{\underline{2}}}

\newcommand{\bm}[1]{\mbox{\boldmath$#1$}}

\def\double #1{#1{\hbox{\kern-2pt $#1$}}}

\newcommand{\gd}{{\dot\g}}

\newif\ifdtup

\newcommand{\bsubeq}{\begin{subequations}}
\newcommand{\esubeq}{\end{subequations}}

\numberwithin{equation}{section}

\newcommand{\sSU}{\mathsf{SU}}

\newcommand{\sU}{\mathsf{U}}

\begin{document}

\begin{titlepage}
\begin{flushright}
January, 2023 \\
Revised version: March, 2023
\end{flushright}
\vspace{5mm}

\begin{center}
{\Large \bf 
On higher-spin $\bm{\cN=2}$ supercurrent multiplets

}
\end{center}

\begin{center}

{\bf Sergei M. Kuzenko and Emmanouil S. N. Raptakis} \\
\vspace{5mm}

\footnotesize{
{\it Department of Physics M013, The University of Western Australia\\
35 Stirling Highway, Perth W.A. 6009, Australia}}  
~\\
\vspace{2mm}
~\\
Email: \texttt{ 
sergei.kuzenko@uwa.edu.au, emmanouil.raptakis@research.uwa.edu.au}\\
\vspace{2mm}

\end{center}

\begin{abstract}
\baselineskip=14pt
We elaborate on the structure of higher-spin $\mathcal{N}=2$ supercurrent multiplets 
in four dimensions. It is shown that associated with every conformal supercurrent 
$J_{\alpha(m) \dot{\alpha}(n)}$ (with $m,n$ non-negative integers) is a descendant 
 $J^{ij}_{\a(m+1) \ad(n+1)}$ with the following properties: (a) it is a linear multiplet with respect to its $\mathsf{SU}(2)$ indices, that is 
 $ D_\beta^{(i}  J^{  jk)}_{\alpha(m+1) \dot{\alpha}(n+1) }=0$ and 
$ \bar D_{\dot \beta}^{(i}  J^{jk)}_{ \alpha(m+1) \dot{\alpha}(n+1)  }=0$;  and  (b) it is conserved, 
  $\partial^{\beta \dot{\beta}} J^{ij}_{\beta \alpha(m) \dot{\beta} \dot{\alpha}(n)}=0$. Realisations of the conformal supercurrents $J_{\alpha(s) \dot{\alpha}(s)}$, with $s=0,1, \dots$, are naturally provided by a massless hypermultiplet and a vector multiplet. It turns out that such supercurrents and their linear descendants $J^{ij}_{\alpha(s+1) \dot{\alpha}(s+1)}$ do not occur in the harmonic-superspace framework recently described by Buchbinder, Ivanov and Zaigraev. 
Making use of a massive hypermultiplet, we derive non-conformal higher-spin 
$\mathcal{N}=2$ supercurrent multiplets. Additionally, we derive the higher symmetries of the kinetic operators for both a massive and massless hypermultiplet. Building on this analysis, we sketch the construction of higher-derivative gauge transformations for the off-shell arctic multiplet $\U^{(1)}$, which are expected to be 
vital in the framework of consistent interactions between $\U^{(1)}$ and superconformal higher-spin gauge multiplets.
\end{abstract}
\vspace{5mm}

\vfill

\vfill
\end{titlepage}

\newpage
\renewcommand{\thefootnote}{\arabic{footnote}}
\setcounter{footnote}{0}

\tableofcontents{}
\vspace{1cm}
\bigskip\hrule

\allowdisplaybreaks

\section{Introduction}

Almost two years ago, we described all possible $\N=2$ conformal supercurrent multiplets 
$J^{\a(m) \ad(n)}$ in a curved supergravity background \cite{KR21}, with $m$ and $n$ non-negative integers. The $m=n=0$ case corresponds to the ordinary conformal supercurrent $J$ 
\cite{Sohnius, HST}, which is the source of the Weyl multiplet of $\cN=2$ conformal supergravity
\cite{BdeRdeW}. The supergravity origin of $J$ and its non-conformal extensions was uncovered in \cite{KT,BK10}. If at least one of the parameters $m,n$ is non-zero, 
the supercurrent $J^{\a(m) \ad(n)}$ is the source of a superconformal primary higher-spin gauge multiplet $\U_{\a(m) \ad(n)}$. 
The corresponding superconformal gauge-invariant action for  
$\U_{\a(m) \ad(n)}$ was constructed in \cite{KR21} in an arbitrary conformally flat background. 

For a massless hypermultiplet, the conserved conformal supercurrent multiplets 
$J^{\a(s) \ad(s)}$, with $s\geq 0$, were derived in \cite{KR21} 
 in an arbitrary conformally flat background. In the $s=0$ case, the conserved massless hypermultiplet supercurrent $J$  exists in an arbitrary supergravity background
 \cite{KT}. It should be stressed that each conserved conformal hypermultiplet supercurrent $J^{\a(s) \ad(s)}$ is uniquely defined by the corresponding conservation equation. In particular, this equation implies that $J^{\a(s) \ad(s)}$ is a conformal primary superfield.

For a massive hypermultiplet in $\cN=2$ Minkowski superspace, conserved  supercurrent multiplets $J^{\a(s) \ad(s)}$ can be derived from the massless ones given in \cite{KR21} by coupling the hypermultiplet to a `frozen' vector multiplet \`a la \cite{BK97}.\footnote{In the $s=0$ case, the conserved massive hypermultiplet supercurrent $J$  exists in an arbitrary supergravity background
 \cite{KT}. The hypermultiplet's mass is generated by coupling it to a vector multiplet, one of the two supergravity compensators. Following \cite{BK-AdS}, the supercurrent conservation equation has the form
 $\frac{1}{4} (\bar \cD^{ij} + 4 \bar S^{ij}) J = \cW T^{ij}$, where 
 $\bar \cD^{ij} = \bar \cD_\ad^{(i} \bar \cD^{j) \ad }$, $\bar S^{ij}$ is one of the torsion tensors,
 $\cW$ is the chiral field strength of the vector multiplet and $T^{ij}$ is the supertrace obeying the constraints  $\cD_\alpha^{(k} T^{ij)} = 0$ and $ \bar\cD_\ad^{(k} T^{ij)} = 0$. In the massless case, $T^{ij}=0$.   
 }
This is achieved by replacing the flat covariant derivatives $D_A =(\pa_a , D_\a^i, \bar D^{\ad}_i)$ with gauge-covariant ones ${\bm D}_A$, eq. \eqref{3.1}, with the chiral field strength  
being  constant.

It should be pointed out that the conserved higher-spin supercurrent multiplets for a free hypermultiplet have also been described in terms of $\cN=1$ superfields. In the massless hypermultiplet case, the corresponding higher-spin conformal supercurrents were derived in \cite{KMT}  in $\cN=1$ Minkowski superspace. In $\cN=1$ anti-de Sitter superspace, the higher-spin hypermultiplet supercurrents were described in \cite{BHK}, both in the massless and massive cases.  

Recently, using  the $\cN=2$ harmonic superspace approach,
conserved hypermultiplet superfield currents have been presented by Buchbinder, Ivanov and Zaigraev \cite{BIZ3},  building on their earlier construction of massless higher-spin $\cN=2$ supersymmetric gauge actions \cite{BIZ1,BIZ2}. No comparison with the existing results on the hypermultiplet supercurrent multiplets \cite{KR21,KMT,BHK} 
was given in \cite{BIZ3}. In our opinion, carrying out such a comparison is important, and this short paper provides a thorough analysis.

There exists a duality between massless higher-spin (super)field actions and non-conformal higher-spin (super)currents. In general, massless models are formulated in terms of a uniquely defined conformal gauge (super)field in conjunction with certain compensator (super)fields. In the non-supersymmetric case, this is true for Fronsdal's 
massless spin-$s$ field action \cite{Fronsdal}, with $s\geq 2$ an integer, which is formulated using two fields, the conformal prepotential $h_{\a(s) \ad(s)}$ and the compensator $h_{\a(s-2) \ad(s-2)}$. This is also true in the $\cN=1$ supersymmetric case. For instance, 
there are two off-shell formulations for the massless half-integer superspin multiplet 
\cite{KPS} which involve the same conformal prepotential $H_{\a(s) \ad(s)}$, which is real and unconstrained, but different compensators. 
They are 
described in terms of 
the following dynamical variables:
\begin{subequations}
\bea
\cV^\bot_{s+1/2}& = &\Big\{H_{\a(s)\ad(s)}~, ~
\G_{\a(s-1) \ad(s-1)}~,
~ \bar{\G}_{\a(s-1) \ad(s-1)} \Big\} ~,    \\
\label{10}
\cV^{\|}_{s+1/2} &=& 
\Big\{H_{\a(s)\ad(s)}~, ~
G_{\a(s-1) \ad(s-1)}~,
~ \bar{G}_{\a(s-1) \ad(s-1)} \Big\}
~.
\eea
\end{subequations}
Here the complex superfields 
$\G_{\a (s-1) \ad (s-1)} $ and 
$G_{\a (s-1) \ad (s-1)}$ are  transverse linear and 
longitudinal linear, respectively,
\begin{subequations}
\bea
{\bar D}^\bd \,\G_{\a(s-1) \bd \ad(s-2)} &=&  0 ~,
\label{transverse}
\\
{\bar D}_{ (\ad_1} \,G_{\a(s-1) \ad_2 \dots \ad_{s})}&=&0 ~.
\label{longitudinal}
\eea
\end{subequations}
Different choices of conformal compensators lead to different non-conformal conservation equations \cite{BHK,Hutomo:2017phh}.
In the $\cN=2$ supersymmetric case, the problem of classifying different off-shell massless higher-superspin models is equivalent to that of classifying non-conformal higher-spin supercurrent multiplets. We believe that the massless higher-spin $\cN=2$ supersymmetric gauge models proposed in \cite{BIZ1} are not unique. One way to uncover other off-shell massless higher-superspin actions is to study different non-conformal higher-spin supercurrent multiplets. This paper initiates the program of 
classifying the non-conformal higher-spin $\cN=2$ supercurrent multiplets.

This paper is organised as follows. In section \ref{section2} we review the conformal $\cN=2$ higher-spin supercurrents and corresponding multiplets in the case of a massless hypermultiplet. Additionally, we construct conformal higher-spin supercurrents for a vector multiplet. Section \ref{section3} is devoted to the construction of non-conformal higher-spin supercurrents for a massive hypermultiplet.
The former are also compared with the supercurrents recently proposed in \cite{BIZ3}. The main body of this paper is accompanied by two technical appendicies. In appendix \ref{appendixA} we review conformal supercurrents in $\cN=1$ superspace. Appendix \ref{appendixB} is devoted to an analysis of the higher symmetries of the hypermultiplet. Higher-derivative gauge transformations for the off-shell arctic multiplet $\U^{(1)}$ are also sketched.


\section{Conformal higher-spin supercurrents} \label{section2}

The conformal  $\N=2$ supercurrent multiplets 
$J^{\a(m) \ad(n)}$ were introduced in \cite{KR21}
in a curved supergravity background. 
For our purposes it suffices to work in $\cN=2$ Minkowski superspace where the algebra of covariant derivatives $D_A =(\pa_a , D_\a^i, \bar D^{\ad}_i)$ is determined by  
\bea
\{ D_\a^i , \bar{D}^\bd_j \} = - 2 \ri \,\d_j^i \pa_\a{}^\bd ~,
\label{2.1}
\eea
with all other graded commutation relations being trivial. 


\subsection{$\cN=2$ conserved supercurrent multiplets}
\label{section2.1}

A conformal primary\footnote{See, e.g. \cite{KT,Park} for the definition of a primary tensor superfield in flat superspace.} tensor superfield $J^{\a(m) \ad(n)}$, with $m,n \geq 1$, is said to be a conformal supercurrent if it obeys the constraints \cite{KR21}
\begin{subequations}
	\label{SC1}
	\bea
	D_\b^i J^{\b \a(m-1) \ad(n)} &=& 0 \quad \Longrightarrow \quad D^{ij} J^{\a(m) \ad(n)} = 0 ~, \\
	\bar{D}_\bd^i J^{\a(m) \bd \ad(n-1)} &=& 0 \quad \Longrightarrow \quad \bar{D}^{ij} J^{\a(m) \ad(n)} = 0 ~.
	\eea
\end{subequations}
Here we have employed the shorthand
\bea
{D}^{ij} = {D}^{\a(i} \bar{D}_\a^{ j)} ~, \qquad
\bar{D}^{ij} = \bar{D}_{\ad}^{(i} \bar{D}^{\ad j)} 
~.
\eea
In the special case $m=n=s$, the supercurrent $J^{\a(s) \ad(s)}$ 
is restricted to be real. This case was first described in \cite{HST}.
Setting $n = 0$, the conformal supercurrents $J^{\a(m)}$ instead satisfy
\begin{subequations}
	\label{SC2}
	\bea
	D_\b^i J^{\b \a(m-1)} &=& 0 \quad \Longrightarrow \quad D^{ij} J^{\a(m) } = 0 ~, \\
	\bar{D}^{ij} J^{\a(m)} &=& 0 ~.
	\eea
\end{subequations}
Finally, when $m=n=0$, $J$ is real and corresponds to the conformal supercurrent of \cite{HST,Sohnius,KT}
\bea
\label{SC3}
D^{ij} J = 0 ~, \quad \bar{D}^{ij} J = 0 ~.
\eea
The above constraints imply that, at the component level,  
$J^{\a(m) \ad(n)}$ contains a number of conserved currents.
The component analysis of $J^{\a(m) \ad(n)}$ can be performed in two stages. 
First, each supercurrent $J^{\a(m) \ad(n)}$ can be decomposed as a collection of 
conformal $\cN=1$ supercurrent multiplets, which  
is discussed in appendix \ref{appendixA}.
Second, every conformal $\cN=1$ supercurrent can be reduced to components, resulting in several ordinary conformal currents. This stage is described in \cite{KR19}. One can find a general local solution to the constraints  \eqref{SC1}, \eqref{SC2} and \eqref{SC3} using the $\cN=2$ superspin projection operators
\cite{Hutchings}.
 
The constraints \eqref{SC1}, \eqref{SC2} and \eqref{SC3} may also be naturally formulated within the analytic superspace\footnote{We refer the reader to \cite{HH} for a detailed review of this superspace formulation} \cite{HL2}. It would be interesting to determine the supercurrents for the hypermultiplet \eqref{HMSC} and vector multiplet \eqref{VMSC} within this formulation.
 
As discussed in \cite{KR21}, conformal Killing tensors may be utilised to construct new conserved supercurrent multiplets from existing ones. We recall that the former are primary tensor superfields $\z_{\a(p) \ad(q)}$, $p,q \geq 0$, satisfying
\bea
\label{SCKTMB}
D_{(\a_1}^i \z_{\a_2 \dots \a_{p+1}) \ad(q)} = 0 ~, \quad \bar D_{(\ad_1}^i \z_{\a(p) \ad_2 \dots \ad_{q+1})} = 0 ~.
\eea
In particular, given a conformal supercurrent $J^{\a(m)\ad(n)}$ and a conformal Killing\footnote{The conformal Killing tensors of Minkowski superspace were introduced in \cite{HL1,HL3}.} 
tensor $\z_{\a(p) \ad(q)}$, with $m \geq p$ and $n \geq q$, it may be shown that
\bea
\label{2.7}
\mathfrak{J}^{\a(m-p) \ad(n-q)} = J^{\a(m-p) \b(p) \ad(n-q) \ad(q)} \z_{\b(p) \bd(q)} ~,
\eea
also constitutes a conformal supercurrent.\footnote{It should be noted that an analogue of \eqref{2.7} holds in analytic superspace, see \cite{HL2} for further details.}


\subsection{The hypermultiplet}

An on-shell massless hypermultiplet is described by a conformal primary isospinor $q^i$ (and its conjugate $\bar q_i$) subject to the constraints
\bea
\label{HM1}
D_{\a}^{(i} q^{j)} = 0~, \quad \bar D_{\ad}^{(i} q^{j)} = 0~.
\eea
From $q^i$ and its conjugate, we construct the higher-derivative descendants \cite{KR21}
\bea
\label{HMSC}
J^{\a(s) \ad(s)} &=& - \frac{\ri^{s}}{2} \sum^s_{k=0} (-1)^k {s \choose k}^2 \pa^{(\a_1 (\ad_1} \dots \pa^{\a_k \ad_k} q^i \pa^{\a_{k+1} \ad_{k+1}} \dots \pa^{\a_s) \ad_s)} \bar{q}_i \non \\
&&+ \frac{\ri^{s+1}}{16} \sum_{k=0}^{s-1} (-1)^k {s \choose k} {s \choose k+1} \non \\
&&\qquad  \times 
\bigg\{ \pa^{(\a_1 (\ad_1} \dots \pa^{\a_k \ad_k} D^{\a_{k+1} i} q_i \pa^{\a_{k+2} \ad_{k+1}} \dots \pa^{\a_s) \ad_{s-1}} \bar D^{\ad_s) j} \bar q_j \non \\
&&\qquad \quad - \pa^{(\a_1 (\ad_1} \dots \pa^{\a_k \ad_k} \bar D^{\ad_{k+1} i} q_i \pa^{\a_{k+1} \ad_{k+2}} \dots \pa^{\a_{s-1} \ad_{s})} D^{\a_s) j}
\bar q_j
\bigg\} ~.
\eea
It may be shown that the superfield \eqref{HMSC} is conformally primary and satisfies the conservation equations \eqref{SC1} for $s>0$ and \eqref{SC3} when $s=0$. Thus, 
$J^{\a(s) \ad(s)}$ is a conformal supercurrent.
The $s=0$ case was studied in a supergravity background 
in \cite{KT}. 


\subsection{The vector multiplet}

Next, we consider an on-shell vector multiplet, which is described by a primary complex scalar $W$ (and its conjugate $\bar W$) subject to the constraints
\bea
\label{2.11}
\bar{D}_{\ad}^{i} W = 0~, \quad D^{ij} W = 0~.
\eea
From $W$ and its conjugate, one may construct the higher-derivative descendants\footnote{It should be noted that this supercurrent may be extended to general conformally-flat backgrounds by the replacement $D_A \rightarrow \nabla_A$, where $\nabla_A$ is the conformally covariant derivative, see e.g. \cite{ButterN=2,KRTM2}.}
\bea
\label{VMSC}
J^{\a(s) \ad(s)} &=& \ri^s \sum^s_{k=0} (-1)^k {s \choose k}^2 \pa^{(\a_1 (\ad_1} \dots \pa^{\a_k \ad_k} W \pa^{\a_{k+1} \ad_{k+1}} \dots \pa^{\a_s) \ad_s)} \bar{W} \non \\
&-& \frac{\ri^{s+1}}{2} \sum_{k=0}^{s-1} (-1)^{k} {s \choose k} {s \choose k+1} 
\pa^{(\a_1 (\ad_1} \dots \pa^{\a_k \ad_k} D^{\a_{k+1} i} W \non \\
&& \qquad \qquad \times \pa^{\a_{k+2} \ad_{k+1}} \dots \pa^{\a_s) \ad_{s-1}} \bar D^{\ad_s)}_i \bar W \non \\
&+& \frac{\ri^{s}}{16} \sum_{k=0}^{s-2} (-1)^{k} {s \choose k} {s \choose k+2} 
\pa^{(\a_1 (\ad_1} \dots \pa^{\a_k \ad_k} D^{\a_{k+1} \a_{k+2}} W \non \\
&& \qquad \qquad \times \pa^{\a_{k+3} \ad_{k+1}} \dots \pa^{\a_s) \ad_{s-2}} \bar D^{\ad_{s-1} \ad_s)} \bar W ~,
\eea
where we have made the definitions
\bea
D_{\a \b} = D_{(\a}^i D_{\b)i} ~, \qquad \bar{D}_{\ad \bd} = \bar{D}_{(\ad i} \bar{D}_{\bd)}^i~.
\eea
The $s=0$ case was considered 
in \cite{HST}. It may be shown that \eqref{VMSC} is primary and satisfies the conservation equations \eqref{SC1} for $s>0$ and \eqref{SC3} when $s=0$, hence it constitutes a conformal supercurrent.
 
 
 \subsection{Linear descendants} 
 \label{section2.4}
 
 Given a conformal supercurrent $J_{\a(m) \ad(n)}$, 
 the conservation equations \eqref{SC1}, \eqref{SC2} and \eqref{SC3} imply the relation
\bea 
\big[ D^{ij} , \bar{D}^{kl} \big] J_{\a(m) \ad(n)} =0~. 
 \eea
 Its use allows one to demonstrate that the following 
 descendant of $J^{\a(m) \ad(n)} $,
 \bea
 J^{ij}_{\a(m+1) \ad(n+1)} 
 := \hf \big[D_{(\a_1}^{(i} , \bar D^{j)}_{(\ad_1}\big] 
 J_{\a_2 \dots \a_{m+1}) \ad_2 \dots \ad_{n+1} )} 
 \label{2.12}
 \eea
 is Grassmann analytic, 
\begin{subequations} \label{2.13}
 \bea
 D_\b^{(i}  J^{  jk)}_{\a(m+1) \ad(n+1) }=0~, \qquad 
 \bar D_\bd^{(i}  J^{jk)}_{ \a(m+1) \ad(n+1)  }=0~, 
\label{2.13a}
 \eea 
 and conserved, 
 \bea
 \pa^{\b\bd} J^{ij}_{\a(m) \b  \ad(n)\bd  }=0~.
 \label{2.13b}
 \eea
 \end{subequations}
 It is not difficult to check that, unlike $J_{\a(m) \ad(n)}$,
 the descendant $J^{ij}_{\a(m+1) \ad(n+1)} $ is not a conformal primary superfield. 
It follows from \eqref{2.13a} that $J^{ij}_{\a(m+1) \ad(n+1)} $ is a linear multiplet with respect to its $\sSU(2)$ indices. 
Using the projective-superspace terminology 
\cite{G-RRWLvU,K2010},  $J^{ij}_{\a(m+1) \ad(n+1)} $ can also be called 
an $\cO(2)$ multiplet with respect to its $\sSU(2)$ indices. 

As is well known  \cite{HST,BS}, given a linear multiplet $G^{ij} = G^{ji}$ constrained by 
 $D_\b^{(i}  G^{  jk)}=0$ and $ \bar D_\bd^{(i}  G^{jk)}=0$, it contains a transverse vector defined by $V_{\b\bd} = \frac{\ri}{2} [D^i_\b, \bar D^j_\bd ] G_{ij}$ such that $\pa^{\b\bd} V_{\b\bd}=0$.
  Therefore, associated with the descendant \eqref{2.12} is the superfield 
 \bea
 J_{\b\bd, \a(m+1) \ad(n+1)} := \frac{\ri}{2} [D_{\b i}, \bar D_{\bd j}] J^{ij}_ {\a(m+1) \ad(n+1)} ~,
 \label{2.144}
 \eea
 which is separately transverse in its $\beta \bd$ and $\a \ad$ indices, 
\begin{subequations} \label{2.155}
 \bea
 \pa^{\b\bd} J_{\b\bd, \a(m+1) \ad(n+1)} &=&0~,  \label{2.155a}\\
 \pa^{\g\gd} J_{\b\bd, \g \a(m) \gd \ad(n)} &=&0~. \label{2.155b}
 \eea
 \end{subequations}
 
For comparison with the results of \cite{BIZ3}, it is convenient to start using the harmonic superspace conventions \cite{GIKOS} (see \cite{GIOS} for a review).
 We introduce harmonics  $u^+_i$ and $u^-_i$ parametrising the group $\sSU(2)$, 
 \bea
 \overline{u^{+i} } = u^-_i ~, \qquad u^{+i} u^-_i =1~,
 \eea 
 and define a new basis for the spinor covariant derivatives, 
 \bea
 D^\pm_\a = D^i_\a \,u^\pm_i~, \qquad
{\bar D}^\pm_\ad = {\bar D}^i_\ad \,u^\pm_i~,
\eea
such that eq. \eqref{2.1}
 takes the form
\bea
 \{ {\bar D}^+_\ad , D^-_\a \}
= - \{ D^+_\a , {\bar D}^-_\ad \}
=2{\rm i}\, \pa_{\a \ad}~, 
\eea
The operators  $D^+_\a$ and $\bar D^+_\ad$ anticommute with each other.
By definition, the analytic superfields $\f^{(n)} (z,u^\pm)$ are annihilated by $D^+_\a$ and $\bar D^+_\ad$.
The superscript `$n$' of  $\f^{(n)} $ denotes the harmonic $\sU(1)$ charge of  $\f^{(n)} $ defined by $D^0  \f^{(n)} =n  \f^{(n)} $,
where $D^0$ is one of 
  the harmonic derivatives 
\bea
D^{++}=u^{+i}\frac{\partial}{\partial u^{- i}} ~,\quad
D^{--}=u^{- i}\frac{\partial}{\partial u^{+ i}} ~,\quad
D^0&=&u^{+i}\frac{\partial}{\partial u^{+i}}-u^{-i}
\frac{\partial}{\partial u^{-i}} ~,
\eea
which form a basis in the space of left-invariant vector fields on $\sSU(2)$.

Associated with the descendant \eqref{2.12} is the harmonic superfield 
\bea
J^{++ \a(m+1) \ad(n+1)} = u^+_i u^+_j  J^{ij \a(m+1) \ad(n+1)} 
\label{2.18}
\eea
with the following properties:
\begin{subequations} \label{2.19}
\bea
&D^+_\b J^{++ \a(m+1) \ad(n+1)} =0~, \qquad \bar D^+_\bd J^{++ \a(m+1) \ad(n+1)} =0~,
\label{2.19a}\\
&D^{++} J^{++ \a(m+1) \ad(n+1)} =0~, \label{2.19b}\\
&\pa_{\b\bd} J^{++ \a(m)\b  \ad(n) \bd} =0~. \label{2.19c}
\eea
\end{subequations}

In the case of the hypermultiplet supercurrent \eqref{HMSC}, its descendant 
$J^{++\a(s+1)\ad(s+1)} $ has the form:
\begin{align}
\label{analyticdescendant}
J^{++\a(s+1)\ad(s+1)} = \ri^{s+1} \sum_{k=0}^{s+1} (-1)^k {s+1 \choose k}^2 \pa^{(\a_1(\ad_1} \dots \pa^{\a_k \ad_k} q^+ \pa^{\a_{k+1} \ad_{k+1}} \dots \pa^{\a_{s+1}) \ad_{s+1})} \breve{q}^{+}~,
\end{align}
where $\breve{q}{}^+ = -{\bar q}^i u^+_i$ is the smile-conjugate of $q^+=q^iu^+_i$. It obeys the Grassmann analyticity constraints \eqref{2.19a}
since the on-shell hypermultiplet superfields $q^+$ and $\breve{q}^+$ are Grassmann analytic. It obeys the constraint \eqref{2.19b} since both $q^+$ and $\breve{q}^+$ are annihilated by $D^{++}$. Finally, the explicit dependence of $J^{++\a(s+1)\ad(s+1)} $ on spacetime derivatives guarantees the fulfilment of \eqref{2.19c}.

Now we reproduce the current superfield for the massless hypermultiplet given in 
\cite{BIZ3}
\begin{align}
J^{++ \a(s+1) \ad(s+1)}_{\rm BIZ} &= - \frac{\ri^{\frac{1 - (-1)^s}{2}}}{2} 
\Big \{ q^{+} \partial^{(\a_1(\ad_1} \dots \partial^{\a_{s+1})\ad_{s+1})} \breve{q}^+
- (-1)^s \breve{q}^{+} \partial^{(\a_1(\ad_1} \dots \partial^{\a_{s+1})\ad_{s+1})} {q}^+
\non \\
&\phantom{=} + \xi (1 - (-1)^s) \partial^{(\a_1(\ad_1} \dots \partial^{\a_{s+1})\ad_{s+1})} (q^+ \breve{q}^+) \Big \} ~,
\label{2.21}
\end{align}
where $\xi$ is a real parameter. It satisfies only the equations \eqref{2.19a} and \eqref{2.19b}. 
Therefore, at the component level it contains a single conserved current.
It is  defined to be the $\q$-independent component of \eqref{2.144} and satisfies only the conservation equation 
\eqref{2.155a}. 

Actually, there are $2 \lfloor \frac{s+3}{2} \rfloor$ linearly independent real 
 superfields of the given tensor structure, 
 ${\mathfrak J}^{++\a(s+1)\ad(s+1)}$,
  that satisfy the same equations \eqref{2.19a} and \eqref{2.19b}. Their general form is
\begin{align}
\label{2.24}
{\mathfrak J}^{++\a(s+1)\ad(s+1)} = 
\sum_{k=0}^{s+1} 
c_k\,
\pa^{(\a_1(\ad_1} \dots \pa^{\a_k \ad_k} q^+ \pa^{\a_{k+1} \ad_{k+1}} \dots \pa^{\a_{s+1}) \ad_{s+1})} \breve{q}^{+} = \breve{\mathfrak J}^{++\a(s+1)\ad(s+1)} 
~,
\end{align}
with arbitrary coefficients $c_k$ modulo natural reality conditions.
It is natural to ask the question:
What is special about the supercurrent \eqref{2.21}?

The supercurrent \eqref{2.21} was derived in \cite{BIZ3} 
via the Noether procedure, which involved making use of the rigid symmetries of the hypermultiplet introduced in \cite{BIZ2}. 
Without requiring the hypermultiplet transformation laws to be compatible with the $\cN=2$ superconformal symmetry, there is a huge freedom in the structure of such rigid 
symmetries of the hypermultiplet and, as a result, in the explicit form of cubic interaction vertices and associated conserved currents. In Refs. \cite{BIZ3,BIZ2}, the authors fixed a particular non-conformal family of rigid symmetries of the hypermultiplet.
Consequently, 
the corresponding conserved currents \eqref{2.21} are non-primary, which is in contrast to the multiplet \eqref{HMSC}.\footnote{We refer the reader to \cite{KPR} for a study of the Noether procedure for matter coupled to (super)conformal higher-spin fields with manifest (super)conformal invariance.} The latter should naturally arise by gauging the higher symmetries \eqref{B5} - \eqref{B.6}, which preserve superconformal symmetry. Other currents \eqref{2.24} correspond to different choices of rigid symmetries of the hypermultiplet.

A general local solution to the equations \eqref{2.19} is given by \eqref{2.18} 
in which $J^{ij \a(m+1) \ad(n+1)} $ has the form 
\begin{subequations} \label{2.23}
\bea
J^{ij}_{ \a(m+1) \ad(n+1)} &=& D^{ij} \F_{\a(m+1) \ad(n+1)} +
\bar D^{ij} \bar \J_{ \a(m+1) \ad(n+1)} ~, \label{2.23a} \\
\bar D_j^\bd  \F_{\a(m+1) \ad(n+1)} &=&0~, \qquad 
\bar D_j^\bd  \J_{\a(n+1) \ad(m+1)} =0~,  \label{2.23b}\\
  \F_{\a(m+1) \ad(n+1)} &=& \pa^{\b_1}{}_{\ad_1} \dots \pa^{\b_{n+1}}{}_{\ad_{n+1}} 
  \pa_{(\b_1}{}^{\bd_1} \dots \pa_{\b_{n+1}}{}^{\bd_{n+1}}
 \f_{\a_1 \dots \a_{m+1}) \bd_1 \dots \bd_{n+1} }~, \label{2.23c}\\
 \J_{\a(n+1) \ad(m+1)} &=& \pa^{\b_1}{}_{\ad_1} \dots \pa^{\b_{m+1}}{}_{\ad_{m+1}} 
  \pa_{(\b_1}{}^{\bd_1} \dots \pa_{\b_{m+1}}{}^{\bd_{m+1}}
 \j_{\a_1 \dots \a_{n+1}) \bd_1 \dots \bd_{m+1} }~,\label{2.23d}
\eea
\end{subequations}
with $\f_{\a(m+1) \ad(n+1)}$ and $  \j_{\a(n+1) \ad(m+1)} $ being arbitrary chiral superfields. 
In the $m=n$ case, these chiral superfields coincide.
The relations \eqref{2.23c} and \eqref{2.23d} are not included if only the constraints \eqref{2.19a} and \eqref{2.19b} are imposed, as in the case of \eqref{2.21}.


\section{Non-conformal higher-spin supercurrents} \label{section3}

This section is devoted to an analysis of non-conformal higher-spin supercurrents for the massive hypermultiplet. Additionally, by making use of the results of the previous subsection, the modified conservation law of the analytic descendant \eqref{analyticdescendant} is derived.

Before turning to the higher-spin story, it is worth recalling the explicit structure of the non-conformal supercurrent \cite{BK-AdS} in the $\cN=2$ supergravity formulation of \cite{deWPV}. 
This supergravity formulation makes use of two compensators:
 the vector multiplet  and the tensor multiplet.
In the superspace setting, the Weyl multiplet is described in terms of the conformal prepotential $\cH $, which is real and unconstrained.  The vector multiplet is described using Mezincescu's prepotential $\cV_{ij}$, which is a real iso-triplet. The prepotentials for the tensor multiplet are a chiral scalar $\J$ and its conjugate. The supercurrent conservation equation, which corresponds to a matter model with action $S$, was derived in \cite{BK-AdS} and has the form
(with $\cW$ and $\cG^{ij}$ being the field strengths of the vector and tensor compensators, respectively) 
\begin{align}
\label{N2current}
\frac{1}{4} (\bar \cD^{ij} + 4 \bar S^{ij}) \cJ = \cW \cT^{ij} - \cG^{ij} \cY~,
\end{align}
where $\cJ$ denotes the supercurrent, while $\cT^{ij}$ and $\cY$ are the trace multiplets. 
\bea
\cJ = \frac{{ \d} S}{{ \d}\cH }~, \qquad \cT^{ij} = \frac{{\d} S}{{\d} \cV_{ij} }~, 
\qquad \cY = \frac{{\d} S}{{\d} \J }~.
\eea
The multiplet $\cJ$ and $\cT_{ij}$ must be real,  and $\cY$ covariantly  chiral.
In addition, $\cY$ and $\cT_{ij}$ must obey the constraints
\begin{subequations}
\begin{gather}\label{eq_CurrentConstraints}
\cD_\a^{(k} \cT^{ij)} = \bar\cD_\ad^{(k} \cT^{ij)} = 0 ~,\\
(\cD^{ij} + 4 S^{ij})\cY = (\bar \cD^{ij} + 4 \bar S^{ij})\bar\cY~.
\end{gather}
\end{subequations}

In the higher-spin case, we know the analogue of $\cH$, which is the superconformal prepotential 
$H_{\a(s) \ad(s)}$, though we do not know yet the higher-spin analogues of the compensators $\cV_{ij}$ and $\J$. We hope to learn about their structure by analysing non-conformal higher-spin supercurrents. 


\subsection{The massive hypermultiplet}
\label{Section3.1}
 
In flat superspace, it is possible to endow 
a hypermultiplet with a mass by coupling it to a `frozen' vector multiplet \`a la \cite{BK97}. This approach is most useful in dealing with off-shell hypermultiplets without intrinsic central charge, such as the $q^+$ hypermultiplet \cite{GIKOS} and the polar hypermultiplet (see \cite{G-RRWLvU,K2010} and references therein).
It can also be used for the Fayet-Sohnius hypermultiplet \cite{Fayet, Sohnius78}
which suffices for our goals. 
To this end, we introduce the gauge covariant derivatives
\bea
\bm{D}_A = (\bm{D}_a,\bm{D}_\a^i,\bar{\bm{D}}^{\ad}_i) = D_A +V_A \D ~,
\label{3.1}
\eea
where $\D$ is a central charge operator, $[\D, \bm{D}_A] = 0$. The covariant derivatives $\bm{D}_A$ are characterised by the anti-commutation relations
\begin{subequations}\label{VMAlgebra}
\bea
&\{ \bm{D}_\a^i , \bm{D}_\b^j \} = - 2 \ve^{ij} \ve_{\a\b}  \bar{\bm{W}}_0 \D ~, 
\quad \{{\bar {\bm D}}_{\dot \a i} \, ,  {\bar {\bm D}}_{\dot  \b j} \} = - 2 
\ve_{ij}\, \ve_{\dot \a \dot \b} {\bm W}_0 \D  ~, 
\\
& \{ \bm{D}_\a^i , \bar{\bm{D}}^\bd_j \} = - 2 \ri \d_j^i \bm{D}_\a{}^\bd ~,
\eea
\end{subequations}
where $\bm{W}_0$ is a constant non-zero parameter.
The gauge freedom may be used to bring the gauge-covariant derivatives to the form
\begin{subequations}\label{CCcovder1}
\bea
{\bm D}_\a^i &=& \phantom{-} \frac{\pa}{\pa \q^{\a}_i}
+ {\rm i} \,(\s^b )_{\a \bd} \, {\bar \q}^{\dot \b i}\, \pa_b
- \q^i_\a \,\bar{\bm W}_0  \D  = D^i_\a - \q^i_\a \,\bar{\bm  W}_0   \D~, 
\\
{\bar {\bm D}}_{\dot \a i} &=&
- \frac{\pa}{\pa {\bar \q}^{\dot \a i}} 
- {\rm i} \, \q^\b _i (\s^b )_{\b \dot \a} \,\pa_b
+{\bar \q}_{\dot \a i} \,  {\bm W}_0  \D = {\bar D}_{\ad i}+ {\bar \q}_{\dot \a i} \,{\bm W}_0   \D ~. ~~~
\eea
\end{subequations}
Without loss of generality, it is possible to choose
\be
\bm{W}_0 = \ri~.
\ee

To describe a massive hypermultiplet,  the constraints \eqref{HM1} should be replaced with
\be
\label{HMVM}
\bm{D}_{\a}^{(i} q^{j)} = 0~, \quad \bar{\bm{D}}_{\ad}^{(i} q^{j)} = 0~.
\ee
An important consequence of \eqref{HMVM} is
\bea
(\bm{D}^a \bm{D}_a + \D^2 ) {q}^i = 0 ~.
\eea
This is just a different way of looking at the Fayet-Sohnius hypermultiplet \cite{Fayet, Sohnius78}.
Thus, requiring that $q^i$ is an eigenvector of $\D$ (with non-zero eigenvalue)
\bea
\D q^i = \ri m q^i ~, \qquad m \in \mathbb{R}\setminus\{0\}~,
\eea
is equivalent to endowing the hypermultiplet with a mass,
\bea
(\bm{D}^a \bm{D}_a - m^2 ) q^i = 0~.
\eea
Indeed, in the realisation \eqref{CCcovder1} we have ${\bm D}_a = \pa_a$.

Having formulated the massive hypermultiplet above, 
its higher-spin supercurrents are derived from \eqref{HMSC} by performing the replacement $D_A \rightarrow \bm{D}_A$. They satisfy the conservation equations
\begin{subequations} \label{3.9}
\bea
s=0:& \qquad {D}^{ij} J = \ri \mathbb{T}^{ij} ~, \label{3.9a}\\
s>0:& \qquad {D}_\b^i J^{\a(s-1) \b \ad(s)} = \ri \bar{{D}}^{(\ad_1}_j T^{ ij \,\a(s-1) \ad_2 \dots \ad_s)} + \bar{{D}}^{(\ad_1 i} S^{\a(s-1) \ad_2 \dots \ad_s)} ~, \label{3.9b}
\eea
\end{subequations}
where we have introduced the real supertrace multiplets\footnote{The supertrace $\mathbb{T}^{ij}$ was first derived in \cite{KT}.}
\begin{subequations}
\begin{align}
\mathbb{T}^{ij} &= 4 m \ri q^{(i} \bar{q}^{j)} ~, \\
T_{\a(s-1) \ad(s-1)}^{ij} &= m \ri^{s} \sum_{k=0}^{s-1} \frac{(-1)^{k+1} (2k-s+1)s(s+1)(s+2)}{6(k+1)(k+2)(s-k+1)(s-k)} {s-1 \choose k}^2  \non \\
& \qquad \qquad \times \bm{D}_{(\a_1 (\ad_1} \dots \bm{D}_{\a_k \ad_k} q^{(i} 
\bm{D}_{\a_{k+1} \ad_{k+1}} \dots \bm{D}_{\a_{s-1}) \ad_{s-1})} \bar{q}^{j)} ~, \\
S_{\a(s-1) \ad(s-1)} &= m \ri^{s-1} \sum_{k=0}^{s-1} \frac{(-1)^{k+1} s (s+1)(s+2)(s+3)}{4(k+1)(k+2)(s-k+1)(s-k)} {s-1 \choose k}^2  \non \\
& \qquad \qquad \times \bm{D}_{(\a_1 (\ad_1} \dots \bm{D}_{\a_k \ad_k} q^i 
\bm{D}_{\a_{k+1} \ad_{k+1}} \dots \bm{D}_{\a_{s-1}) \ad_{s-1})} \bar{q}_i ~.
\end{align}
\end{subequations}
These multiplets are neutral with respect to the central charge. They satisfy the following differential constraints:
\begin{subequations} \label{3.11}
\begin{align}
{D}_\b^{(i} \mathbb{T}^{jk)} &= 0 ~, \qquad \bar{{D}}_\bd^{(i} \mathbb{T}^{jk)} = 0~,\label{3.11a} \\
{D}_\b^{(i} T_{\a(s-1) \ad(s-1)}^{ jk)} &= 0 ~, \qquad 
\bar{{D}}_\bd^{(i}  T_{\a(s-1) \ad(s-1)}^{ jk)} = 0~, 
\label{3.11b} \\
{D}^{ij} S_{\a(s-1) \ad(s-1)} &= \bar{{D}}^{ij} S_{\a(s-1) \ad(s-1)} 
~.
\label{3.11c}
\end{align}
\end{subequations}
The latter implies that 
\bea
S^{ij}_{\a (s-1) \ad(s-1) } := 
\frac{1}{4} D^{ij} S_{\a(s-1) \ad(s-1)} 
= \frac 14 \bar{D}^{ij} S_{\a(s-1) \ad(s-1)} 
\eea
is a linear multiplet with respect to its $\sSU(2)$ indices
\bea
{D}_\b^{(i} S_{\a(s-1) \ad(s-1)}^{ jk)} &= 0 ~, 
\qquad 
\bar{{D}}_\bd^{(i}  S_{\a(s-1) \ad(s-1)}^{ jk)} = 0~.
\label{3.133}
\eea

It should be emphasised that there might be a different functional form for the right-hand side of \eqref{3.9b}. This will be discussed elsewhere.


\subsection{Linear descendants}

Let us analyse the massive hypermultiplet supercurrents introduced in the previous subsection. 
 As in the massless case, we can introduce the descendant $ J_{ij}^{\a(s+1) \ad(s+1)}$, which is defined by the rule \eqref{2.12}, and the associated harmonic superfield $J^{++ \a(s+1)\ad(s+1)}$,  eq. \eqref{2.18}. 
 \begin{align}
	J^{++\a(s+1)\ad(s+1)} = \ri^{s+1} \sum_{k=0}^{s+1} (-1)^k {s+1 \choose k}^2 \bm{D}^{(\a_1(\ad_1} \dots \bm{D}^{\a_k \ad_k} q^+ \bm{D}^{\a_{k+1} \ad_{k+1}} \dots \bm{D}^{\a_{s+1}) \ad_{s+1})} \breve{q}^{+}~,
\end{align} 
which is obtained from \eqref{analyticdescendant} upon the replacement $\partial_a \rightarrow \bm{D}_a$.
 Making use of the conservation equations \eqref{3.9} and \eqref{3.11}, 
it is seen that $J^{++ \a(s+1)\ad(s+1)}$ still obeys the constraints \eqref{2.19a} and \eqref{2.19b}. However, the constraint \eqref{2.19c} does not hold anymore. It is replaced with 
\begin{align}
\label{3.12}
\pa_{\bb} J^{++ \a(s)\b\ad(s)\bd} = \frac{s}{s+1}  
\pa^{(\a_1 (\ad_1}S^{++\a_2 \dots \a{s})  \ad_2 \dots \ad_s)}~.
\end{align}
The relations \eqref{3.11a} and \eqref{3.11b} tell us that the harmonic superfields 
$\mathbb{T}^{++} := u^+_iu^+_j \mathbb{T}^{ij}  $ and $T^{++\a(s-1) \ad(s-1)} := u^+_i u^+_j T^{\a(s-1) \ad(s-1) ij}  $ are analytic. For $s>0$ the trace multiplets $S^{\a(s-1) \ad(s-1)} $
and  $T^{++\a(s-1) \ad(s-1)} $ are related to each other as follows:
\bea
D^+_\b \bar D^{+ (\ad_1} S^{\a (s-2)  \b \ad_2 \dots \ad_{s-1}) } 
+ 3 \pa_\b{}^{(\ad_1} T^{++\a (s-2 ) \b  \ad_2 \dots \ad_{s-1})} =0~.
\eea

These results should also be compared with those for the massive hypermultiplet given in 
\cite{BIZ3}. In particular, we expect that they belong to the family of superfields
obtained from \eqref{2.24} upon the replacement $\partial_a \rightarrow \bm{D}_a$
\begin{align}
	{\mathfrak J}^{++\a(s+1)\ad(s+1)} = 
	\sum_{k=0}^{s+1} 
	c_k\,
	\bm{D}^{(\a_1(\ad_1} \dots \bm{D}^{\a_k \ad_k} q^+ \bm{D}^{\a_{k+1} \ad_{k+1}} \dots \bm{D}^{\a_{s+1}) \ad_{s+1})} \breve{q}^{+} = \breve{\mathfrak J}^{++\a(s+1)\ad(s+1)} 
	~.
\end{align}
It is clear that these multiplets obey the constraints \eqref{2.19a} and \eqref{2.19b}, though they do not satisfy an analogue
of \eqref{3.12}. Hence, it is not clear what is special about the non-conformal supercurrents employed in \cite{BIZ3}.


\noindent
{\bf Acknowledgements:}
We are grateful to Michael Ponds for useful discussions.
The work of SK is supported in part by the Australian 
Research Council, project No. DP200101944.
The work of ER is supported by the Hackett Postgraduate Scholarship UWA,
under the Australian Government Research Training Program. 

\appendix

\section{Conformal supercurrents in $\cN=1$ superspace}
\label{appendixA}

This appendix is aimed at reviewing the structure of conformal supercurrents in $\cN=1$ superspace. As shown below, they naturally arise from the $\cN=2$ multiplets of section \ref{section2.1} upon performing a superspace reduction. 

First, we sketch the procedure of reducing from $\cN=2$ to $\cN=1$ Minkowski superspace. Let $D_\a $, $\bar{D}^\ad $ and $\pa_{\aa} = \frac{\ri}{2} \{D_\a , \bar D_\ad \}$ be the covariant derivatives of $\mathbb{M}^{4|4}$. They may be defined via the $\cN=2$ covariant derivatives as follows: $ D_\a {\mathfrak U} =  D_\a^{\underline{1}} U|$
and $\bar{D}^\ad {\mathfrak U}= \bar{D}^{\ad}_{\underline{1}} U|$.
Here $U$ is an $\cN=2$ superfield, and  ${\mathfrak U} \equiv U| :=U|_{\theta^\a_{\underline 2} = \bar{\theta}_\ad^{\underline 2} = 0}$ is its $\N=1$ projection.

We now review the basic properties of $\N = 1$ conserved current supermultiplets\footnote{We refer the reader to \cite{KR19} for a more in-depth analysis.}. A primary tensor superfield $\mathcal J^{\a(m) \ad(n)}$, with $m,n \geq 1$, subject to the constraints
\begin{subequations}\label{supercurrent} 
	\bea
	D_{\b} \mathcal J^{\b\a (m-1)\ad(n)} &=& 0 \quad \implies \quad D^2 \mathcal J^{\a(m) \ad(n)} = 0 ~, 
	\label{supercurrent-a} 
	\\
	\bar D_{\bd} \mathcal J^{\a(m) \bd \ad(n-1) } &=& 0 \quad \implies \quad \bar D^2 \mathcal J^{\a(m) \ad(n)} = 0 ~,
	\label{supercurrent-b} 
	\eea
\end{subequations}
is a conformal supercurrent. 
The $m=n=1$ case corresponds to the ordinary 
conformal supercurrent \cite{FZ}. 
For $m >n =0$, \eqref{supercurrent} should be replaced with
\begin{subequations}\label{supercurrent2} 
	\bea
	D_{\b} \mathcal J^{\b\a (m-1)} &=& 0 \quad \implies \quad D^2 \mathcal J^{\a(m)} = 0 ~, 
	\label{supercurrent2-a} 
	\\
	\bar D^2 \mathcal J^{\a(m)  } &=& 0~.
	\label{supercurrent2-b} 
	\eea
\end{subequations}
The  $m=1$ case was first considered in \cite{KT}, where it was shown that the spinor supercurrent $\cJ^\a$ naturally originates from the reduction of the conformal $\cN=2$ supercurrent \cite{Sohnius}
to $\cN=1$ superspace. 
Finally, when $m=0$ the supercurrent satisfies
\bea
\label{supercurrent3} 
D^2 \mathcal J =0~, \quad 
\bar D^2 \mathcal J = 0~.
\eea
This is the flavour current supermultiplet \cite{FWZ}. The above $\cN=1$ supercurrents each lead to a multiplet of conserved currents at the component level, see \cite{KR19} for a detailed study.

We first consider the $\N=2$ supercurrent $J^{\a(m)\ad(n)}$ \eqref{SC1}. It contains four independent $\N=1$ conserved current supermultiplets
\begin{subequations}
	\label{2.16}
	\bea
	j^{\a(m) \ad(n)} &=& J^{\a(m) \ad(n)} | ~, \\
	j^{\a(m+1) \ad(n)} &=& D^{(\a_1 \underline 2} J^{\a_2 \dots \a_{m+1}) \ad(n)} | ~, \\
	j^{\a(m) \ad(n+1)} &=& \bar D^{(\ad_1}_{\underline 2} J^{\a(m) \ad_2 \dots \ad_{n+1})} | ~, \\
	j^{\a(m+1) \ad(n+1)} &=& \hf \big[ D^{ (\a_1 \underline 2} , \bar D_{\underline{2}}^{(\ad_1} \big] J^{\a_2 \dots \a_{m+1}) \ad_2 \dots \ad_{n+1})} | \non \\
	&-& \frac{1}{2(m+n+3)} \big[ D^{(\a_1} , \bar D^{(\ad_1} \big] j^{\a_2 \dots \a_{m+1}) \ad_2 \dots \ad_{n+1})} \non \\
	&-& \frac{\ri (m-n)}{m+n+3} D^{(\a_1 (\ad_1} j^{\a_2 \dots \a_{m+1}) \ad_2 \dots \ad_{n+1})} ~.
	\eea
\end{subequations}
Similarly, $J^{\a(m)}$ \eqref{SC2} is composed of four $\N=1$ supercurrents
\begin{subequations}
	\bea
	j^{\a(m)} &=& J^{\a(m)} | ~, \\
	j^{\a(m+1)} &=& D^{(\a_1 \underline 2} J^{\a_2 \dots \a_{m+1})} | ~, \\
	j^{\a(m) \ad} &=& \bar D^{\ad}_{\underline 2} J^{\a(m)} | ~, \\
	j^{\a(m+1) \ad} &=& \hf \big[ D^{(\a_1 \underline 2} , \bar D_{\underline{2}}^{\ad} \big] J^{\a_2 \dots \a_{m+1})} | - \frac{1}{2(m+3)} \big[ D^{(\a_1} , \bar D^{\ad} \big] j^{\a_2 \dots \a_{m+1})} \non \\
	&-& \frac{\ri m}{m+3} \pa^{(\a_1 \ad} j^{\a_2 \dots \a_{m+1})} ~.
	\eea
\end{subequations}
Finally, upon reduction of $J$ \eqref{SC3} we obtain three $\N=1$ current multiplets \cite{KT}
\begin{subequations}
	\bea
	j^{} &=& J^{} | ~, \\
	j^{\a} &=& D^{\a \underline 2} J | ~, \\
	j^{\a \ad} &=& \hf \big[ D^{\a \underline 2} , \bar D_{\underline{2}}^{\ad} \big] J | - \frac{1}{6} \big[ D^{\a} , \bar D^{\ad} \big] j ~.
	\eea
\end{subequations}

\section{Higher symmetries of the hypermultiplet}
\label{appendixB}

This appendix is devoted to the study of the higher symmetries of the on-shell
hypermultiplet in four-dimensional $\N = 2$ Minkowski superspace.\footnote{An off-shell extension of our analysis is also sketched.} This analysis is sequential to that of \cite{KLRTM}, where such studies were undertaken for the massless hypermultiplet in a background of off-shell $\N = (1,0)$ supergravity in six spacetime dimensions. Unlike \cite{KLRTM}, however, in four dimensions it is possible to endow this supermultiplet with a mass. We begin by studying the massless case and then proceed to examine the massive case, which breaks superconformal symmetry, allowing us to derive the Killing condition for tensor superfields. As an extension of this analysis, we also sketch the construction of higher-derivative gauge transformations for the off-shell arctic multiplet $\U^{(1)}$. 

\subsection{The massless hypermultiplet}

We recall that an on-shell massless hypermultiplet is described by the primary isospinor $q^i$ (and its conjugate $\bar q_i$) subject to the conformally-invariant constraints \eqref{HM1}. A scalar linear differential operator $\mathfrak D$ is said to be a symmetry of the massless hypermultiplet if it preserves the superconformal properties of $q^i$ 
in addition to the constraints \eqref{HM1}
\bea
\label{B.2}
D_\a^{(i} {\mathfrak D} q^{j)} = 0 ~, \qquad \bar D_\ad^{(i} {\mathfrak D} q^{j)} = 0~.
\eea
Two symmetry operators ${\mathfrak D}$ and $\widetilde {\mathfrak D}$ are said to be equivalent, 
$ {\mathfrak D} \sim \widetilde{\mathfrak D}$, provided
\bea
\label{B.3}
{\mathfrak D} \sim \widetilde{\mathfrak D} \quad \Longleftrightarrow \quad  (\mathfrak{D} - \widetilde{\mathfrak D}) q^i = 0 
~.
\eea

Given a positive integer $n$, we look for an $n$th-order symmetry operator 
\begin{subequations}
	\bea
	{\mathfrak D}_\xi^{[n]} 
	& = & \sum_{k=0}^{n} \xi^{A_{1} \dots A_{k}} D_{A_{k}} \dots D_{A_{1}}  +  \sum_{k=0}^{n-1} \xi^{A_{1} \dots A_{k} j(2)} D_{A_{k}} \dots D_{A_{1}} J_{j(2)} ~,
	\eea
	where $J_{ij}$ is the $\sSU(2)_R$ generator\footnote{We recall that $J_{ij}$ acts on a generic isospinor $\chi_k$ by the rule $J_{ij} \chi_k = - \ve_{k(i} \chi_{j)}$.} and the coefficients may be chosen to be graded symmetric 
	\bea
	\xi^{A_{1} \dots A_{i} A_{i+1} \dots A_{k}} = (-1)^{\ve_{A_{i}} \ve_{A_{i+1}}} \xi^{A_{1} \dots A_{i+1} A_{i} \dots A_{k}} ~, \qquad 1\leq i \leq k-1~, \\
	\xi^{A_{1} \dots A_{i} A_{i+1} \dots A_{k} j(2)} = (-1)^{\ve_{A_{i}} \ve_{A_{i+1}}} \xi^{A_{1} \dots A_{i+1} A_{i} \dots A_{k} j(2)} ~, \qquad 1\leq i \leq k-1~.
	\eea
\end{subequations}
Modulo the equivalence \eqref{B.3},
${\mathfrak D}_\xi^{[n]} $ may be brought to a canonical form given by
\bea
\label{B5}
	\mathfrak{D}_\xi^{[n]} q^{i} &=& \Big \{ \sum_{k=0}^{n} \xi^{\a(k) \ad(k)} (\pa_{\aa})^k + \sum_{k=0}^{n-1} \Big \{ \xi^{\a(k+1) \ad(k)}{}_j (\pa_{\aa})^k D_{\a}^j + \xi^{\a(k) \ad(k+1)}{}_j (\pa_{\aa})^k \bar{D}_{\ad}^j \non \\
	&\phantom{=}& + \xi^{\a(k) \ad(k) j(2)} (\pa_{\aa})^k J_{j(2)} \Big \} \Big \} q^{i}
\eea

We now impose the conditions \eqref{B.2} and require that $\mathfrak{D}^{[n]}_\xi$ preserves all superconformal properties of $q^i$. This allows us to determine all coefficients appearing in \eqref{B5} in terms of $\xi^{\a(n) \ad(n)}$, which proves to be a conformal Killing tensor superfield \eqref{SCKTMB}
\begin{align}
	\label{SCKT}
	D_{(\a_1}^i \xi_{\a_2 \dots \a_{n+1}) \ad(n)} = 0~, \qquad \bar{D}_{(\ad_1}^i \xi_{\a(n) \ad_2 \dots \ad_{n+1})} = 0~.
\end{align}
Specifically, we find:
\begin{subequations}
	\label{B.6}
\bea
	\xi^{\a(k) \ad(k)} &=&  \bigg [ \frac{n((n+k+2)(k+1)+n(n-k))}{2(n+1)^3} \binom{n+k}{2k} \binom{2k}{k} \binom{2n}{n-1}^{-1}(\pa_\bb)^{n-k} \non \\
	&\phantom{=}& + \frac{(-1)^k (n-1)^2}{32(n+1)^2(2n-1)} \binom{n+k}{k} \binom{n-2}{k} \binom{2n-2}{n}^{-1} (\pa_\bb)^{n-k-2} \big \{ D_{\b(2)} , \bar{D}_{\bd(2)} \big \} \bigg ] \non \\
	&\phantom{=}& \times \xi^{\a(k) \b(n-k) \ad(k) \bd(n-k)} ~, \\
	\xi^{\a(k+1) \ad(k) i} &=& \bigg [ \frac{\ri n^2(k+2)}{2(n+1)(k+1)} \binom{n+k+1}{k} \binom{n-1}{k} \binom{2n}{n-1}^{-1} (\pa_\bb)^{n-k-1} \bar{D}_\bd^i \non \\
	&\phantom{=}& - \frac{4(-1)^k(n+k+1)}{k+1} \binom{n+k+1}{k} \binom{n-2}{k} \binom{2n-2}{n}^{-1} (\pa_\bb)^{n-k-2} D_\b^i \bar{D}_{\bd(2)} \bigg ] \non \\
	&\phantom{=}& \times \xi^{\a(k+1) \b(n-k-1) \ad(k) \bd(n-k)} ~, \\
	\xi^{\a(k) \ad(k+1) i} &=& \bigg [ - \frac{\ri n^2(k+2)}{2(n+1)(k+1)} \binom{n+k+1}{k} \binom{n-1}{k} \binom{2n}{n-1}^{-1} (\pa_\bb)^{n-k-1} D_\b^i \non \\
	&\phantom{=}& - \frac{4(-1)^k(n+k+1)}{k+1} \binom{n+k+1}{k} \binom{n-2}{k} \binom{2n-2}{n}^{-1} (\pa_\bb)^{n-k-2} \bar{D}_\bd^i {D}_{\b(2)} \bigg ] \non \\
	&\phantom{=}& \times \xi^{\a(k) \b(n-k) \ad(k+1) \bd(n-k-1)} ~, \\
	\xi^{\a(k) \ad(k) j(2)} &=& \bigg [ \frac{\ri n^2}{2(n+1)^2} \binom{n+k+1}{k} \binom{n-1}{k} \binom{2n}{n-1}^{-1} (\pa_\bb)^{n-k-1} D_\b^{(j_1} \bar{D}_\bd^{j_2)} \bigg ] \non \\
	&\phantom{=}& \times \xi^{\a(k) \b(n-k) \ad(k) \bd(n-k)} ~.
\eea
\end{subequations}

Thus, the higher symmetries of the massless hypermultiplet in flat superspace take the form \eqref{B5}, where the coefficients are given by the expressions \eqref{B.6}. As a result, such symmetries are determined in terms of a single superfield parameter, a conformal Killing tensor, subject to the constraints \eqref{SCKT}.

The above analysis admits an extension to the off-shell polar hypermultiplet
(following the projective-superspace terminology \cite{KRTM2, LR2, G-RRWLvU}).
To derive it, we make use of an isotwistor $v^i \in \mathbb{C}^2 \setminus \{0\}$, defined modulo the equivalence relation $v^i \sim \mathfrak{c} v^i$, where $\mathfrak{c} \in \mathbb{C} \setminus \{0\}$, hence it constitutes inhomogeneous coordinates for $\mathbb{C}P^1$. In projective superspace, the on-shell polar hypermultiplet is described by the index-free homogeneous polynomial 
\bea
q^{(1)} (v)= v_i q^i
\eea
 obeying the analyticity constraints
\bea
D_\a^{(1)} q^{(1)} = v_i D_\a^i q^{(1)} = 0~, \qquad \bar{D}_\ad^{(1)} q^{(1)} = v_i \bar{D}_\ad^i q^{(1)}=0~.
\eea
To recast \eqref{B5} in terms of $q^{(1)}$, it is useful to introduce a new isotwistor $u^i$ subject to the constraint $(v,u) := v^i u_i \neq 0$. The result is as follows:
\bea
	\label{B5new}
	{\mathfrak D}_\xi^{[n]} q^{(1)} &=& \Big \{ \sum_{k=0}^{n} \xi^{\a(k) \ad(k)} (\pa_{\aa})^k - \sum_{k=0}^{n-1} \Big \{ \xi^{\a(k+1) \ad(k) (1)} {D}_{\a}^{(-1)}
	+ \xi^{\a(k) \ad(k+1) (1)} (\pa_{\aa})^k \bar{D}_{\ad}^{(-1)} \Big \} \non \\
	&\phantom{=}& - \sum_{k=0}^{n-1} \Big \{ \xi^{\a(k) \ad(k) (2)} (\pa_{\aa})^k \pa^{(-2)} - \xi^{\a(k) \ad(k) (0)} (\pa_{\aa})^k \Big \} \Big \} q^{(1)} ~,
\eea
where the parameters appearing in \eqref{B5new} are related to those of \eqref{B5} by the rules
\begin{subequations}\label{B.10}
\bea 
	\xi^{\a(k+1)\ad(k) (1)} &=& v_j \xi^{\a(k+1)\ad(k) j} ~, \qquad	~~\xi^{\a(k)\ad(k+1) (1)} = v_j \xi^{\a(k)\ad(k+1) j} ~,  \\
	\xi^{\a(k)\ad(k) (2)} &=&  v_{j_1} v_{j_2} \xi^{\a(k)\ad(k) j_1 j_2} ~, \qquad \xi^{\a(k)\ad(k) (0)} = \frac{v_{j_1} u_{j_2}}{(v,u)} \xi^{\a(k)\ad(k) j_1 j_2} ~,
\eea
\end{subequations}
and we have defined the differential operators
\begin{subequations}
\bea
	D_\a^{(-1)} &=& \frac{u_i}{(v,u)} D_\a^i \qquad \bar{D}_\ad^{(-1)} = \frac{u_i}{(v,u)} \bar{D}_\ad^i ~, \\
	\pa^{(-2)} &=& \frac{1}{(v,u)} u^i \frac{\pa}{\pa v^i} ~.
\eea
\end{subequations}
It should be emphasised that obtaining \eqref{B5new} from \eqref{B5} did not require use of the on-shell condition \eqref{HM1}. 

An off-shell polar hypermultiplet is described in terms of an arctic weight-1 multiplet 
$\U^{(1)}(v)$ and its smile-conjugate antarctic multiplet $\breve{\U}^{(1)} (v)$, which are defined in the north and south charts of ${\mathbb C}P^1 $, respectively.
By definition, the {\it north chart} of ${\mathbb C}P^1$ consists of those points for which 
the first component  of $v^i = (v^{\1}, v^{\2})$ is non-zero,  $v^{\1} \neq 0$.
The north chart of ${\mathbb C}P^1$ may be parametrised by the inhomogeneous complex coordinate 
$\z= v^{\2}/v^{\1} \in \mathbb C$. The only point of ${\mathbb C}P^1 $ outside the north 
chart is characterised by $v_\infty^i = (0, v^{\2})$ and describes an infinitely separated point.
Thus one may think of the projective space ${\mathbb C}P^1 $ as  
the Riemann sphere
${\mathbb C} \cup \{\infty \}$. 
The {\it south chart} of ${\mathbb C}P^1$ is defined to consist of those points for which 
the second component  of $v^i = (v^{\1}, v^{\2})$ is non-zero,  $v^{\2} \neq 0$.
The south chart is naturally parametrized by $1/\z$.
The intersection of the north and south charts is ${\mathbb C} \setminus \{ 0\}$. 

By definition, the off-shell arctic weight-$1$ multiplet, $\U^{(1)} (v)$, is  
holomorphic  in the north chart of ${\mathbb C}P^1$
\bea
\U^{(1)} ( v) &=&  v^{\1}\, \U ( \z) ~, \qquad 
\U ( \z) = \sum_{k=0}^{\infty} \U_k  \z^k 
~. 
\label{arctic1}
\eea
Its smile-conjugate {\it antarctic} multiplet, $\breve{\U}^{(1)} (v) $, has the explicit form 
 \bea
\breve{\U}^{(1)} (v) &=& v^{\2}  \, \breve{\U}(\z) =
v^{\1} \,\z \, \breve{\U} (\z) ~, \quad
\breve{\U}( \z) = \sum_{k=0}^{\infty}  {\bar \U}_k \,
\frac{(-1)^k}{\z^k}
\label{antarctic1}
\eea
and is holomorphic in the south chart of ${\mathbb C}P^1$. 
Setting $\U^{(1)} ( v) = q^{(1)} ( v)$ gives the on-shell hypermultiplet. 

Our crucial observation is that the transformation 
\bea
\d \U^{(1)} ={\mathfrak D}_\xi^{[n]} \U^{(1)} ~,
\eea
with the operator ${\mathfrak D}_\xi^{[n]} $ being given by the relations 
\eqref{SCKT}, \eqref{B.6}, 
\eqref{B5new} and \eqref{B.10}, preserves all the properties of the arctic multiplet. 
It will be shown elsewhere that this transformation is a symmetry of the massless hypermultiplet action 
provided $\x_{\a(n) \ad(n)}$ is subject to a reality condition. We remind the reader that 
this action has the form 
\bea
S=  \frac{\ri}{2\p} \oint_{\g}  (v, \rd v)
\int {\rm d}^4x \, D^{(-4)} \big( \breve{\U}^{(1)} \U^{(1)} \big) |_{\q_i = \bar {\q}^i = 0}
~,
\label{PAP}
\eea
where $\g$ denotes a closed integration contour, 
and $D^{(-4)}$ is the  fourth-order differential operator:
\bea
D^{(-4)} := \frac{1}{16} (D^{(-1) })^2  (\bar D^{(-1)})^2~.
\eea
As is obvious, ${\mathfrak D}_\xi^{[0]}$ is a symmetry of \eqref{PAP} if $\x$ is an imaginary number. The first-order operator ${\mathfrak D}_\xi^{[1]} $ is a symmetry of \eqref{PAP} if $\x^{\a\ad} $ is real \cite{K07}.

It is also of interest to analyse possible off-shell gauge transformation rules for $\U^{(1)}$ which are consistent with its kinematic properties. They will allow us, in a future work, to construct a gauge-invariant action for $\U^{(1)}$ coupled to an infinite tower of $\cN=2$ superconformal higher-spin gauge multiplets.\footnote{In \cite{KPR} the Noether procedure was utilised to couple $\cN=0$ and $\cN=1$ matter to background (super)conformal higher-spin multiplets.} Such transformations must preserve the analyticity constraints
\begin{align}
	\label{B.17}
	D_\a^{(1)} \U^{(1)} = 0 ~, \qquad \bar{D}_\ad^{(1)} \U^{(1)} = 0~,
\end{align}
in addition to the superconformal properties and $u$-independence of $\U^{(1)}$.

Here we will restrict our attention to infinitesimal variations of $\U^{(1)}$ taking the form:
\begin{align}
	\label{B.18}
	\d_{\O}^{[s]} \U^{(1)} = D^{(4)} \Big \{ \O^{(-2)\a(s) \ad(s)} (\pa_\aa)^s \pa^{(-2)}  + \dots \Big \} \U^{(1)} ~, \quad s \geq 0~,
\end{align}
where $\O^{(-2)\a(s) \ad(s)}$ is a primary isotwistor superfield of weight $-(s+2)$, the ellipses denote additional terms necessary to preserve the off-shell properties of $\U^{(1)}$ and $D^{(4)}$ is the fourth-order differential operator:
\bea
D^{(4)} := \frac{1}{16} (D^{(1) })^2  (\bar D^{(1)})^2~.
\eea
Requiring $\d_{\O}^{[s]} \U^{(1)} $ to be primary, 
the complete structure of \eqref{B.18} 
can be determined
for general $s$ and this solution will be given elsewhere. Instead, we spell out the two simplest cases:
\begin{subequations}
\begin{align}
	\d_\O^{[0]} \U^{(1)} &= D^{(4)} \Big \{ \O^{(-2)} \pa^{(-2)}  + \hf \pa^{(-2)} \O^{(-2)} \Big \} \U^{(1)} ~, \label{B.20a} \\
	\d_\O^{[1]} \U^{(1)} &= D^{(4)} \Big \{ \O^{(-2) \aa} \pa_{\aa} \pa^{(-2)}  - \frac{\ri}{2} \O^{(-2) \aa} D_\a^{(-1)} \bar{D}_\ad^{(-1)} - \frac{\ri}{2} \bar{D}_\ad^{(1)} \O^{(-2) \aa} \pa^{(-2)} D_\a^{(-1)} \non \\
	&\qquad \quad + \frac{\ri}{2} {D}_\a^{(1)} \O^{(-2) \aa} \pa^{(-2)} \bar{D}_\ad^{(-1)} - \frac{\ri}{4} D_\a^{(1)} \bar{D}_{\ad}^{(1)} \O^{(-2) \aa} \pa^{(-2)} \pa^{(-2)} \Big \} \U^{(1)} ~.
\end{align}
\end{subequations}
Gauge transformation law \eqref{B.20a} was introduced for the first time in the context of five-dimensional $\cN=1$ supergravity \cite{KT-M2007}.


\subsection{The massive hypermultiplet}
\label{AppendixB.2}

To conclude, we briefly examine the higher symmetries of a massive hypermultiplet. We recall from section \ref{Section3.1} that the hypermultiplet may be endowed with a mass by coupling it to a `frozen' vector multiplet. To do this, we require the gauge covariant derivatives $\bm{D}_A$, which were defined in \eqref{3.1} and satisfy the algebra \eqref{VMAlgebra}. The hypermultiplet then satisfies the massive equations of motion \eqref{HMVM}, which break superconformal symmetry.

As the gauge group now includes central charge transformations generated by $\D$, the higher symmetries \eqref{B5} should be altered. It turns out that the necessary modification is as follows:
\begin{align}
	\label{3.85}
	\bold{\hat{\mathfrak{D}}}^{[n]}_\xi = \mathfrak{D}^{[n]}_\xi + \sum_{k=0}^{n-1} \chi^{\a(k) \ad(k)} (\bm{D}_{\aa})^k \D~,
\end{align}
for some tensor superfields $\chi^{\a(k)\ad(k)}$. Imposing the conditions
\begin{align}
	\bm{D}_\a^{(i} \bold{\hat{\mathfrak{D}}}_\xi^{[n]} q^{j)} = 0~, \qquad \bar{\bm{D}}_\ad^{(i} \bold{\hat{\mathfrak{D}}}_\xi^{[n]} q^{j)} = 0~,
\end{align}
we obtain the constraints
\begin{align}
	\label{3.87}
	{\bm{D}}^{\b i} \chi^{\a(k) \ad(k)} = 2 \ri \xi^{\b \a(k) \ad(k) i} ~.
\end{align}
We emphasise that this implies that $\chi^{\a(k) \ad(k)}$ is, in general, a non-local function of $\xi^{\a(n) \ad(n)}$.
Equation \eqref{3.87} implies the following non-trivial condition
\begin{align}
	\label{B.8}
	{\bm{D}}^i_\b \bar{\bm{D}}_{\bd i} \xi^{\a(n-1) \b \ad(n-1) \bd} = 0 \quad \implies \quad \bm{D}_\bb \xi^{\a(n-1) \b \ad(n-1) \bd} = 0~.
\end{align}

To conclude, we note that the imposition of \eqref{B.8} significantly simplifies the structure of the higher symmetries \eqref{B5} -- \eqref{B.6}. In particular, one finds:
\bea
\label{B.9}
{\mathfrak D}_\xi^{[n]} q^i &=&  \Big \{\xi^{\a(n) \ad(n)} (\bm{D}_{\aa})^n + \frac{\ri n}{2(n+1)}\bar{\bm{D}}_{\bd j}\xi^{\a(n) \ad(n-1) \bd} (\bm{D}_{\aa})^{n-1} \bm{D}_{\a}^j
\non \\
&\phantom{=}&
- \frac{\ri n}{2(n+1)} \bm{D}_{\b j}\xi^{\a(n-1) \b \ad(n)} (\bm{D}_{\aa})^{n-1} \bar{\bm D}_{\ad}^j \non \\
&\phantom{=}&
+ \frac{\ri n^2}{2(n+1)^2} \bm{D}_\b^{j_1} \bar{\bm{D}}_\bd^{j_2} \xi^{\a(n-1) \b \ad(n-1) \bd} (\bm{{D}}_{\aa})^{n-1} J_{j(2)} \non \\
&\phantom{=}&
+ \chi^{\a(s-1) \ad(s-1)}  (\bm{D}_{\aa})^{s-1} \D \Big \} q^i ~.
\eea


\begin{footnotesize}

\end{footnotesize}

\end{document}